\documentstyle[amssymb,aps,preprint]{revtex}

\tightenlines

\begin{document}
\title{Critical Current Density of YBa$_{2}$Cu$_{3}$O$_{7-\delta }$ Low-Angle Grain
Boundaries in Self-Field}
\author{D. T. Verebelyi, C. Cantoni, J. D. Budai and D. K. Christen}
\address{Oak Ridge National Laboratory, Oak Ridge, TN 37831-6061}
\author{H. J. Kim and J. R. Thompson}
\address{Department of Physics, University of Tennessee, Knoxville, TN 37996-1200 and}
\address{Oak Ridge National Laboratory, Oak Ridge, TN 37831-6061}
\date{18 September 2000}
\maketitle
\pacs{74.72.B,74.60.J,74.62.D,74.25.H}
\draft

\begin{abstract}
A study has been perfomed on the superconducting critical current density $%
J_{c}$ flowing across low angle grain boundaries in epitaxial thin films of
YBa$_{2}$Cu$_{3}$O$_{7-\delta }$ (YBCO). \ The materials studied were dual
grain boundary rings deposited on SrTiO$_{3}$ and containing 2$^{\text{o}},$
3$^{\text{o}},$ 5$^{\text{o}}$ and 7$^{\text{o}}$ tilt boundaries. \ The
current density in self-field was determined by magnetometric methods at
temperatures from 5 K to $T_{c}.$ We conclude that at the higher
temperatures of coated conductor applications, there is limited potential
for improving $J_{c}$ by reducing the grain boundary angle below\ $\sim $ 3$%
^{\text{o}}.$

PACS numbers (1999 scheme): \ 74.60.Jg, 74.50.+s, \ 74.76.Bz, \ 74.60.Ge
\end{abstract}

\pagebreak
\narrowtext%
%

Well-aligned grains are necessary for optimum critical current density $%
J_{c} $ in high-temperature superconductors. \ A large misalignment of
adjacent grains suppresses the order parameter at the grain boundary (GB)
and results in Josephson tunneling behavior\cite{Cha60} and a
near-exponential decrease in $J_{c}$ with misorientation angle\cite
{Iva59,Hei69,Hil173}. \ In the case of YBa$_{2}$Cu$_{3}$O$_{7-\delta }$
(YBCO) low-angle grain boundaries\ ($\theta \lesssim $ 4$^{\text{o}}$), the
boundary is comprised of$_{{}}$ a periodic array of dislocation cores
separated by only slightly perturbed material\cite{Chi59} providing a strong
conduction channel wider than the in-plane coherence length. \ This
low-angle regime is particularly important to understanding transport in the
highly-textured coated conductors. Both of the most prominent coated
conductor methodologies, rolling-assisted biaxially textured substrate
(RABiTS) \cite{Goy69,Nor274} and\ ion-beam assisted deposition (IBAD)\cite
{Wu67}, can provide this level of texture. Our previous transport results
concluded that a 2$^{\text{o}}$ GB was inherently different from a 5$^{\text{%
o}}$ GB due to the weak-linked characteristics of the 5$^{\text{o}}$ GB\cite
{Ver76}. \ In that study, grain-like voltage-current relations were
observed, and no reduction in $J_{c}$ could be measured across the 2$^{\text{%
o}}$ GB compared with its adjacent grains. \ However, those results were
regarded as inconclusive since the estimated grain boundary voltages were
less than those of the grain for reasonable sample geometry. \ Due to the
important implications for coated conductors and the fundamentals of
epitaxial YBa$_{2}$Cu$_{3}$O$_{7-\delta }$ (YBCO) films, we have
investigated the properties of low-angle grain boundaries utilizing a high
sensitivity magnetometer technique. From their magnetic moments, we deduce
the persistent currents flowing in thin YBCO rings with and without 2$^{%
\text{o}},$ 3$^{\text{o}}$, 5$^{\text{o}}$, and 7$^{\text{o}}$ [001]-tilt
boundaries, and compare these results. \ 

Films were prepared by pulsed laser deposition\cite{Nor274} of YBCO on SrTiO$%
_{3}$ (STO) bicrystal substrates with a single [001]-tilt boundary. \ Using
standard optical photolithography techniques, rings were produced with an
outside diameter of 3 mm and a width of 100 $\mu $m. \ A ring was patterned
across the GB of the bicrystal, resulting in two GB's included in each ''GB
ring.'' \ For comparison, a companion ring was patterned beyond the
grain-boundary region of each substrate to provide a control sample with the
intragrain properties, which will be referred to as the ''grain ring.'' \
The nominal thickness of the films was 200 nm.

Magnetic measurements were conducted with a Quantum Design MPMS-7
magnetometer. \ The separated, individual ring samples on rectangular STO
substrates were attached to Si mounting disks using cellulose nitrate cement
and placed into a Mylar tube for support. The magnetic field $H$, applied
perpendicular to the plane of the ring, was ramped to 3000 G at 5 K to
induce the maximum circulating persistent currents within the ring. \ The
applied field was then reduced to zero, inducing maximum currents in the
opposite sense. The resulting magnetic moment was measured while increasing
the temperature from 5 K to 95 K in 1 K steps. \ For the grain rings, $J_{c}$
was calculated using the modified Bean critical state model,

$J_{c}$ = $\frac{30\text{ }m_{ring}}{RV-R^{^{\prime }}V^{^{\prime }}}$, \
\qquad \qquad \qquad \qquad \qquad \qquad \qquad \qquad \qquad \qquad \qquad
\qquad Eq.1

where $m_{ring}$ is the magnetic moment in units of emu, $V$ is the volume
of a cylinder with the outside radius $R$ of the ring, and $V^{\prime }$ is
the volume of the ''missing'' cylinder with inside radius $R^{\prime }$. \
For the GB ring, we determined the current across the boundary from the
measured magnetic moment by removing the background signal arising from
currents circulating within the ''strip'' of YBCO but not crossing the GB. \
(This procedure was verified by measurements on an open circuit ring in
which a line was etched across its entire 100 $\mu $m width; from these
data, the Bean model, with a different geometrical factor, also yields the
grain $J_{c}$ of a homogeneous YBCO strip.) \ \ In Table I, the relative
magnitudes of the corrections are illustated by the ratio of the apparent GB 
$J_{c}$ (as deduced from the Bean relation Eq. 1) to the actual $J_{c}$ of
the boundary, all at 5 K.

We first consider and discuss some surprising findings on the lowest angle
grain boundary ( 2$^{\text{o}}$), then return to the measurements on the
remaining low angle GB materials. \ The present results, obtained by
magnetometry of the 2$^{\text{o}}$ GB ring, confirmed our previous transport
measurements by showing no discernible reduction in $J_{c}$ when compared
with the companion grain ring. \ Figure 1 shows the temperature dependence
of $J_{c}$ for two sets of samples with 2$^{\text{o}}$ GB and their
companion grain rings. \ High resolution x-ray diffraction quantified the
exact\ YBCO misorientations: 1.83$^{\text{o}}$ [001]-tilt and 0.13$^{\text{o}%
}$ [100]-tilt. \ From the Franck formula, the calculated dislocation spacing
for a 2$^{\text{o}}$ GB is 11 nm, with a maximum strong channel span of 10
nm approximated from visual inspection of transmission electron micrograph
images. \ This corresponds to a minimum reduction of only 10\% in effective
GB\ length, which is within the expected accuracy range of the data
considering geometrical and experimental errors. \ A possible conclusion is
that the dislocation cores along the GB are no more than 1 nm in size with
respect to suppression of the order parameter. \ 

This simplistic view, however, may be overshadowed by the intrinsic
structure of YBCO films on a cubic substrate. The twin structure of YBCO
could provide an explanation for undiminished $J_{c}$ values across a 2$^{%
\text{o}}$ GB$.$ Twin boundaries are formed in YBCO films at the
tetragonal-to-orthorhombic structural transition upon cooling in oxygen from
high temperature. \ Variations in twin structure are dependent on strain,
ambient oxygen pressure and cooling rate\cite{Spe37}.\ Twin boundaries in
films are found to be preferentially aligned along the $\left\langle
110\right\rangle $ and $\left\langle 1\overline{1}0\right\rangle $ of
lattice-matched cubic substrates. \ In this case, the orthorhombic nature of
the lattice is accommodated by a $\sim $ 1.8$^{\text{o}}$ misalignment of
the (1$\overline{1}$0) planes on either side of a (110) twin boundary\cite
{Bud58}. \ Thus a YBCO film on a SrTiO$_{3}(001)\sin $gle crystal actually
consists of domains with 4 distinct, symmetry-related in-plane orientations.
\ Figure 2 shows a x-ray phi scan through the YBCO (225) peak including both
sides of a 5$^{\text{o}}$ bicrystal. \ Each group of three peaks represent
one grain of the bicrystal. \ The strong central peaks at 0$^{\text{o}}$ and
5.1$^{\text{o}}$ arise from the two twin domains for which the YBCO \{110\}
planes are aligned with the substrate [$HH0$] direction. The difference of
5.1$^{\text{o}}$ between these peaks corresponds to the [001]-axis bicrystal
tilt misorientation. \ The smaller peaks located at $\pm 0.9^{o}$ (as given
by [2 tan$^{-1}(b/a)-90^{o}]$ ) from the central peaks are reflections from
the two possible twin domains associated with the twin planes aligned with
the [$H\overline{H}0$] substrate direction. \ This is also true for the
adjacent grain across the GB. \ We note that although the x-ray analysis
does not describe the morphology of domains, it does reveal\ that twinned
YBCO films on single crystal substrates are populated with possible GB
misorientation angles of 0.9$^{\text{o}},$ 89.1$^{\text{o}},$ and 90$^{\text{%
o}}.$ \ Thus the YBCO boundaries located above the intersection of a
bicrystal substrate with a tilt boundary of $\theta ^{o}$ are not composed
of simply a single well-defined GB of angle $\theta .$ \ Instead, different
YBCO domains can impinge with a variety of misorientation angles, including $%
\theta \pm 0.9^{o}$ and $\theta \pm 89.1^{o}.$ \ It is important to note
that twin boundaries supply no dislocations, while the interface between the
suggested twin domains would produce dislocations with an expected 12 nm
period. \ Such a dilute array of dislocations easily accommodates a
superconducting coherence length between dislocations, resulting in the
typically observed strongly non-linear, well coupled Abrikosov
superconductor. \ In this view, the addition of a single 2$^{\text{o}}$ GB
to the matrix would not be expected to provide additional distinguishable
dissipation.

Let us now consider more generally the critical current density for grain
boundaries with slightly larger angles. \ In transport studies, prominant
tunneling behavior appears to dominate transport chacteristics for a GB
angle above 3-4$^{\text{o}}$, where the grain-like channel between the
dislocation cores becomes smaller than a coherence length. \ The present
magnetometry study on GB rings of 2$^{\text{o}},$ 3$^{\text{o}},$ 5$^{\text{o%
}},$ and 7$^{\text{o}}$ provides precise results for the
temperature-dependent $J_{c}$ in the regime of low angle boundaries. \
Figure 3 shows the temperature dependence of $J_{c}$ for 3$^{\text{o}},$ 5$^{%
\text{o}}$, and 7$^{\text{o}}$ GB rings, compared to their companion grain
rings. \ Also, Table I lists the $J_{c}$ values at 5 K for each
misorientation angle.\ \ \ The 5$^{\text{o}}$ and 7$^{\text{o}}$ tilt GB's
manifest their tunneling conduction in transport studies via a linear
differential $V-J$ characteristic; \ in these magnetometry studies, a
signature of tunneling is a complex in-field $J_{c}$ with a pronounced
dependence on the magnetic field history. \ In contrast, transport studies
of the 2$^{\text{o}}$ GB found grain-like, power law characteristics
indicative of strong, tunneling-free conduction. \ The magnetometry studies
of both the 2$^{\text{o}}$ and 3$^{\text{o}}$ rings revealed in-field $J_{c}$%
's with a qualitatively simple, monotonic falloff with $H$ and with
minimal-to-no dependence on field history. \ These results indicate then, a
robust current conduction for GB's up to 3$^{\text{o}}$, but pronounced
tunneling behavior for GB angles of 5$^{\text{o}}$ and above. \ For the 3$^{%
\text{o}}$ GB, the $J_{c}$ lies a factor of 2-3 below the grain value, which
can be accounted for in part by a reduced cross-section of ''good'' material
by dislocations.

Qualitatively, Fig. 3 shows a systematic reduction of GB current density
with angle. \ The inset in Fig. 3 shows quantitatively the angle-dependence
of $J_{c}$ at temperatures $T$ = 5, 64, and 77 K. \ At low temperatures, the 
$J_{c}$ rises rapidly as the GB angle decreases in this regime of low
angles. For comparison, the (calculated) depairing current density at $T$=0
is included in the inset. \ At the higher temperatures of coated conductor
applications, there is limited potential for improving $J_{c}$ by reducing
the grain boundary angle below\ $\sim$ 3$^{\text{o}}.$ \ The present
materials, in the temperature range of LN$_{2}$, already possess the
technologically desirable current densities 
\mbox{$>$}%
10$^{6}$ A/cm$^{2}$. \ \ \ In this low angle range, the intragrain $J_{c}$
becomes the limiting factor, and should become the focus of further
improvements, especially in the presence of magnetic fields. \ 

We thank D.M. Feldmann and D.C. Larbalestier for scientific discussions and
for providing the 3$^{o}$ substrate. \ \ \ Research co-sponsored by the DOE
Division of Materials Sciences, the DOE Office of Energy Efficiency and
Renewable Energy, Power Technologies, under contract DE-AC05-00OR22725 with
the Oak Ridge National Laboratory, managed by UT-Battelle, LLC.\bigskip 

\newpage\ \ \ \ \ \ \ \ \ \ \ \ \ \ \ \ \ \ \ \ \ \ \ \ \ FIGURE\ CAPTIONS

Fig 1. Magnetometer measurements of persistent current density vs.
temperature for \ two 1.8$^{\text{o}}$ grain boundary rings and their
companion grain rings. Transport data from a 1.8$^{\text{o}}$ single grain
boundary are shown for comparison.

Fig. 2. X-ray $\phi$ scan though the YBCO \{225\} peak on a 5.1$^{\text{o}}$
bicrystal showing the in-plane angular relation between various YBCO
domains. \ 

Fig 3. Magnetometer measurements of critical current density vs. temperature
for 2.8$^{\text{o}}$, 5.1$^{\text{o}}$ and 7$^{\text{o}}$ grain boundary
rings and their companion grain rings from the same deposition and
substrate. \ Inset: \ $J_{c}$ \ versus grain boundary angle at $T$ = 5, 64,
and 77 K, in self field.

\bigskip

\bigskip 
=========================================

Table I. Comparison of critical current density at 5 K for YBCO grain
boundaries and their companion grains. \ Data are calculated values of $%
J_{c} $ from the maximum magnetic moment of persistent currents in a narrow
ring of YBCO on SrTiO$_{3}$ substrates, in self field. \bigskip

\bigskip

\begin{tabular}{ccccc}
Misorientation & GB Ring & Grain Ring & ($J_{c}^{GB}$ / $J_{c}^{Grain})$ & \
\ \ ($J_{c}^{GB,corr}$ / $J_{c}^{GB,Bean})$ \\ 
(degrees) & $J_{c}$ (MA/cm$^{2}$) & $J_{c}$ (MA/cm$^{2}$) &  &  \\ 
\hline\hline
1.8 & 35 & 39 & 0.90 & 1 \\ 
1.8 & 36 & 37 & 0.97 & 1 \\ \hline
2.8 & 9.1 & 19$^{a}$ & 0.48 & 1 \\ 
2.8 & 11.5 & 23 & 0.50 & 0.99 \\ \hline
5.1 & 0.95 & 18 & 0.053 & 0.48 \\ 
5.1 & 1.4 & 34 & 0.041 & 0.56 \\ \hline
7 & 0.85 & 31 & 0.027 & 0.47
\end{tabular}

\bigskip $^{a}$ Value measured with an ''open circuit'' ring in strip
geometry.


\newif\ifabfull\abfulltrue

\bigskip

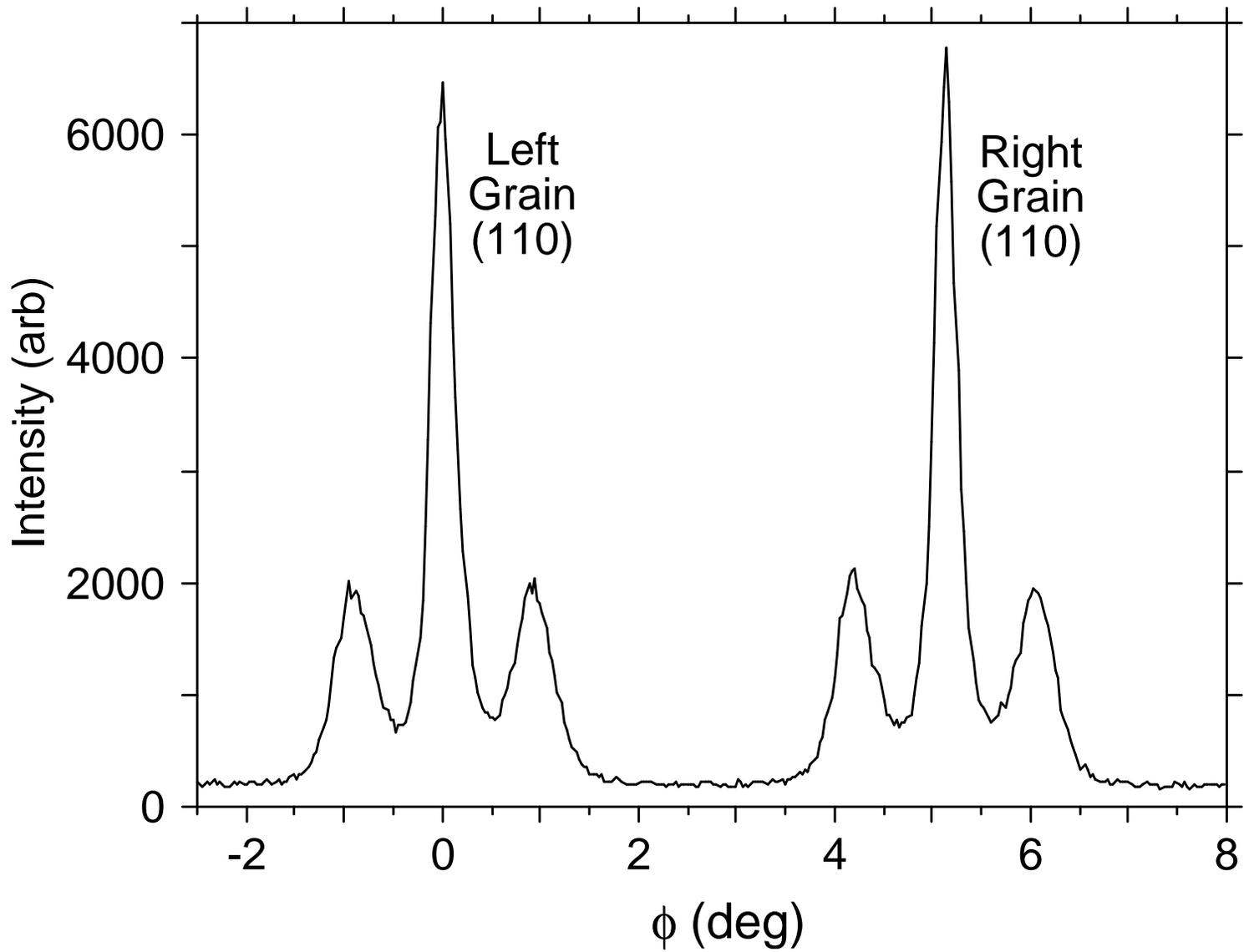
\end{document}